\documentclass{elsart}
\usepackage{latexsym,texdraw,epsfig}
\usepackage{amsfonts,amssymb,amsmath, amstext}

\newcommand{\mev}{\,\mathrm{Me\kern-0.1em V}}
\newcommand{\gev}{\,\mathrm{Ge\kern-0.1em V}}
\newcommand{\re}{\,\mathrm{{\sf I}\kern-0.1em {\sf R}}}
\newcommand{\Tr}{\mathop{\mathrm{Tr}}}

\begin{document}
\begin{frontmatter}
\begin{flushright}
IFUP-TH/2000-32
\end{flushright}
\title{Monopoles, vortices and confinement in $SU(3)$ gauge theory}
\author[Pisa]{L. Del Debbio\thanksref{lddmail}},
\author[Pisa]{A. Di Giacomo\thanksref{adgmail}},
\author[Oxford]{B. Lucini\thanksref{blmail}}

\thanks[lddmail]{ldd@df.unipi.it}
\thanks[adgmail]{digiaco@df.unipi.it}
\thanks[blmail]{lucini@thphys.ox.ac.uk}

\address[Pisa] {Dipartimento di Fisica, Universit\`a di Pisa,
and INFN Sezione di Pisa, Italy}
\address[Oxford] {Theoretical Physics, University of Oxford, UK}

\begin{abstract}
  We compute, in $SU(3)$ pure gauge theory, the vacuum expectation value (vev)
  of the operator which creates a $Z_3$ vortex wrapping the lattice through
  periodic boundary conditions (dual Polyakov line). The technique used is the
  same already tested in the $SU(2)$ case. The dual Polyakov line proves to be
  a good disorder parameter for confinement, and has a similar behaviour to
  the monopole condensate. The new features which characterise the
  construction  of the disorder operator in $SU(3)$  are emphasised.
\end{abstract}

\end{frontmatter}

\setcounter{footnote}{0}

\section{Introduction}

A creation operator $\mu(C)$ of $Z_2$ vortices was recently introduced
for the $SU(2)$ pure Yang-Mills theory~\cite{ldd00}, in order to
investigate their relevance for confinement. We recall that, for a
generic $SU(N)$ theory, $\mu(C)$ creates a defect along the line $C$;
it is defined by the following algebraic relation with the Wilson loop
$W(C^\prime)$~\cite{thooft78}:
\begin{equation}
\label{eq:commut}
\mu(C) W(C^\prime) = W(C^\prime) \mu(C) \exp \left( 
\frac{i 2 \pi n_{CC^\prime}}{N} \right)
\end{equation}
$n_{CC^\prime}$ being the linking number of the curves $C$ and
$C^\prime$. As discussed in~\cite{thooft78}, two operators satisfying
Eq.~\ref{eq:commut} are dual to each other: in the phase characterised by an
area law for $W$, $\mu$ has a perimeter behaviour and viceversa. 

$\mu(C)$ can be viewed as a {\it dual parallel transport}\ along the
line $C$. In what follows we shall choose for $C$ a straight line, say
along the $z$ direction, closing at infinity through periodic boundary
conditions (PBC). $\mu(C)$ on this line can be considered as a {\it
dual Polyakov line}. In Ref.~\cite{ldd00} the behaviour of $\langle
\mu(C) \rangle$ for this line was studied across the $SU(2)$
deconfining phase transition, using techniques introduced in
Refs.~\cite{ldd95,adg97,adg99} to study monopole condensation. The
result was that $\langle \mu(C) \rangle$ is a good disorder parameter
for the deconfinement phase transition, and a finite size scaling
analysis allowed a satisfactory determination of the critical
temperature and indices. In addition it was found that $\langle \mu(C)
\rangle$ coincides within errors with the disorder parameter describing
monopole condensation.

In this paper, we perform a similar analysis for the $SU(3)$ pure gauge
theory, and again we determine the critical temperature and the critical
exponents $\nu$ and $\delta$. As in previous works, instead of $\langle \mu(C)
\rangle$, we study 
\begin{equation}
  \rho = \frac{\partial}{\partial \beta} \log \langle \mu \rangle
\end{equation}
which is more convenient from the numerical point of view, and provides the
same information as $\langle \mu(C) \rangle$. Again we compare with the
analogous quantities defined by magnetic charge condensation, finding
a similar behaviour.

The existing literature on the subject~\cite{kovacs00,rebbi00,delia00}
deals with the $SU(2)$ case and was already compared to our approach
in~\cite{ldd00}.

\section{From $SU(2)$ to $SU(3)$}
\label{sect:vco}

The definition of the vortex creation operator $\mu$ presented in~\cite{ldd00}
can be extended to the case of a $Z_3$ vortex in a straightforward
manner. The peculiarities of this new definition are summarised in
this section, while the reader is referred to~\cite{ldd00} for a
detailed discussion of the method.

The expectation value of $\mu$ in an $SU(N)$ gauge theory is defined
as the ratio of two partition functions:
\begin{equation}
\langle \mu(t_0,x_0,y_0) \rangle = \frac{\tilde Z}{Z} = 
Z^{-1} \int [dU] e^{-\beta \,\tilde S[U]}
\end{equation}
where $Z$ is the usual partition function with the Wilson action,
\begin{equation}
S[U] = \frac{1}{N} \sum_{x,\mu\nu} \Tr \mathrm{Re }\left[1 - P_{\mu\nu}(x)\right]
\end{equation} 
and $\tilde S$ is obtained from $S$ by multiplying a line of plaquettes
in the $0y$ plane by an element of the center:
\begin{equation}
\label{eq:twist}
P_{0y}(t_0, x>x_0, y_0, z) \mapsto \mathbf{z} \, P_{0y}(t_0, x>x_0, y_0, z),
~\forall z
\end{equation}
where $\mathbf{z} \in Z_N$. 

For the $SU(3)$ case discussed in this paper
\begin{equation}
\label{eq:zeds}
\mathbf{z} = \exp\left\{ i\, n \, 2\pi/3 \right\}, \quad \mathrm{with } \quad n=0,1,2 \ ,
\end{equation}
corresponding to the case of zero, one vortex and one antivortex.
Following~\cite{ldd00}, it can be shown by suitable changes of
variables in the definition of $\tilde Z$ that such a prescription
amounts to create a vortex string closed by PBC in the $z$ direction
at $(x_0,y_0,t_0)$, together with an antivortex line at $(N_s-1,y_0,t_0)$
due to PBC in the $x$ direction.

It is worthwhile emphasizing that the vortex creation operator defined
here does not depend on a specific gauge choice. At finite
temperature, it is related to the increase of the free energy when a
vortex is produced in the vacuum of the
theory~\cite{kovacs00,rebbi00,delia00,hart00}. It is gauge-invariant by
construction and relies neither on the identification of the vortices
locations nor on their density nor on any choice of the gauge.

Similarly to the $SU(2)$ case, the measurement of a single operator
$\langle\mu\rangle$ is needed at finite temperature, with $C^*$
boundary conditions to ensure that the net effect of $\mu$ is the
creation of a single vortex as explained in~\cite{adg99}. The free
energy of the vortex configuration is given by:
\begin{equation}
\langle \mu \rangle \propto \exp\left\{ - F/T \right\}
\end{equation}

In pure gauge theory, the confined phase is characterised by the
area law for the Wilson loop, corresponding to a linearly rising
potential between static quarks. With a compact time-dimension, the 
potential can also be determined from the correlation of two Polyakov lines
according to:
\begin{equation}
\label{eq:polcorr}
\langle \mathrm{Tr }~P(R) \: \mathrm{Tr }~P^\dagger(0) \rangle \propto e^{-V(R)
  N_t}
\end{equation}
where $N_t$ is the lattice extension in the compact dimension.
Eq.~\ref{eq:polcorr} and cluster property imply that the vev of a single
Polyakov line has to vanish in the confined phase, that is whenever the Wilson
loop has an area law. One can argue that the dual loops exhibit a similar
behaviour: a non-zero value for their vev corresponds to the confined phase,
characterised by a perimeter law for the dual loop, while a vanishing vev for
the dual Polyakov line corresponds to the phase where the dual loop has an
area law (i.e. the deconfined phase). Perturbative calculations at high
temperature show that indeed $\langle\mu\rangle$ has an area law in this
phase \cite{ak0}, implying vanishing vev for the dual Polyakov line. In the
following section, we shall present non-perturbative lattice results for
$\langle\mu\rangle$ as a function of the temperature.

\section{Numerical results}
\label{sect:num}

In this paper, we present data only for straight vortex lines wrapping in
the $z$ direction by PBC. 

Due to the exponential in its definition, $\langle\mu\rangle$ has large
fluctuations, which make a direct measurement of its expectation value
a challenging task. As usual~\cite{ldd00,ldd95,adg97,adg99}, we focus on the
quantity:
\begin{equation}
\rho = \frac{\partial}{\partial \beta} \log \langle \mu \left( t_0, x_0, y_0
\right) \rangle = \langle S \rangle _S - \langle \tilde{S} \rangle _{\tilde{S}}
\end{equation}
$\rho$ is easy to compute and yields all
the relevant information. At finite temperature, $\rho$ is expected to
have a sharp negative peak in the critical
region~\cite{ldd95,adg97,adg99}, if $\langle \mu \left( t_0, x_0, y_0)
\right \rangle$ is a disorder operator for the deconfinement phase
transition.

\begin{figure}[htp]
\begin{center}
\epsfig{figure=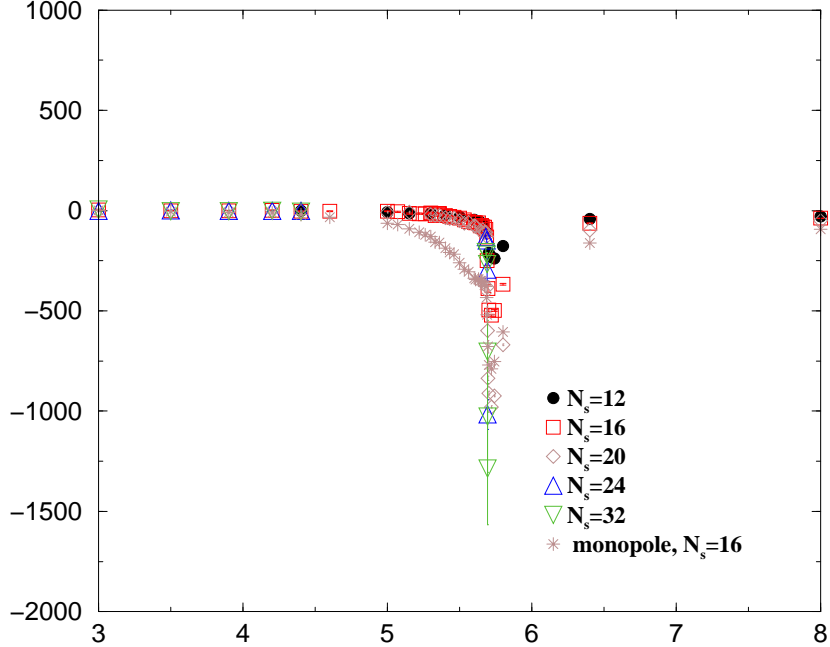, width=11cm} 
\end{center}
\caption{$\rho$ vs. $\beta$ for the $n=1$ vortex on $N_s^3 \times 4$
  lattices. The analogous result for monopoles\cite{adg99} is superimposed for
  comparison.}
\label{fig:finitet1}
\end{figure}

\begin{figure}[htp]
\begin{center}
\epsfig{figure=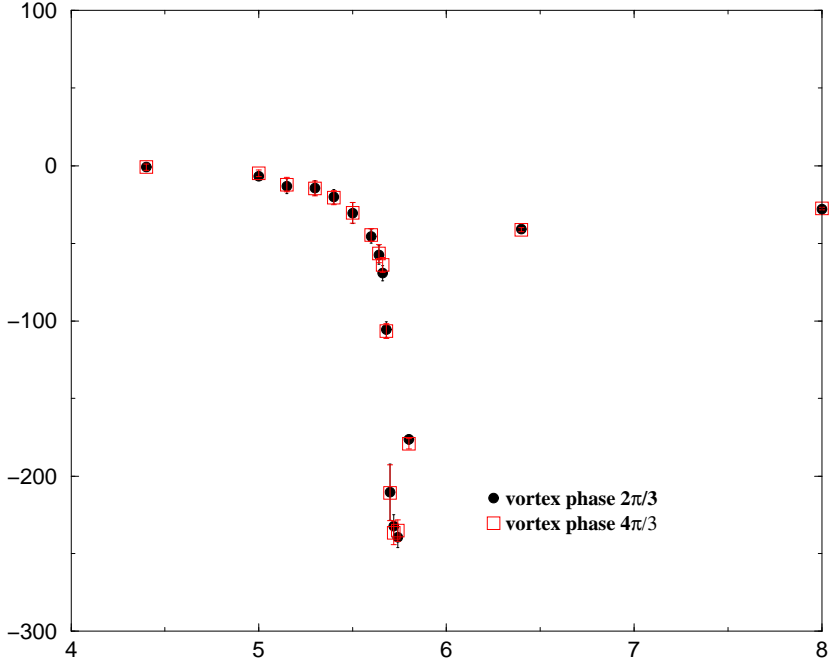, angle=270, width=11cm} 
\end{center}
\caption{Comparison of $\rho$ defined for a vortex and an antivortex
  configuration.}  
\label{fig:figcheck}
\end{figure}

Our results for $\rho$ are displayed in Fig.~\ref{fig:finitet1}, for
several lattice sizes. Fig.~\ref{fig:finitet1} shows the
characteristic peak at the critical coupling, already observed for
$SU(2)$. As a check of our code, we have verified on a $12^3\times 4$
lattice, that $\rho$ has the same behaviour for a vortex and an
antivortex, as expected due to the invariance under charge
conjugation.  The results are shown in Fig.~\ref{fig:figcheck}.

\begin{figure}[htp]
\begin{center}
\epsfig{figure=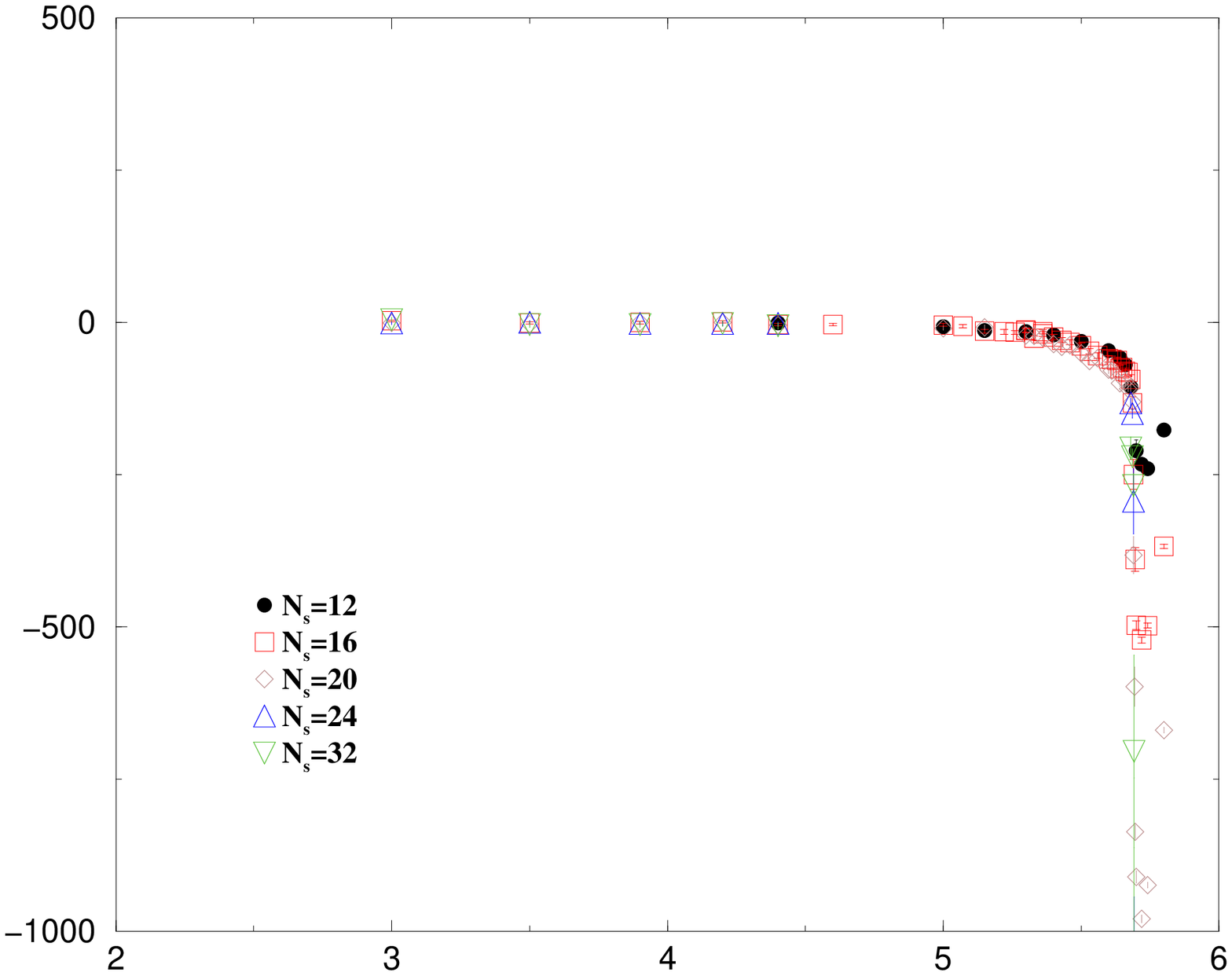, angle=270, width=11cm} 
\end{center}
\caption{Small $\beta$ ehaviour of $\rho$.}  
\label{fig:strong}
\end{figure}

The results for $SU(3)$ are qualitatively very similar to the ones
already presented for $SU(2)$ in~\cite{ldd00}. The value of $\rho$ in
the deconfined phase tends to $-\infty$ in the thermodynamic limit, so
that $\langle \mu \rangle$ vanishes in that limit, in agreement with
the result obtained using perturbation theory in \cite{ak0}. The
small $\beta$ behaviour of $\rho$ is reported in
Fig.~\ref{fig:strong}. Since no change in $\rho$ is observed as the
volume goes large compared to the physical correlation length, we
conclude that in the confined phase $\langle \mu \rangle$ has a a
non-vanishing vev in the thermodynamic limit.

\begin{figure}[htbp]
\begin{center}
\epsfig{figure=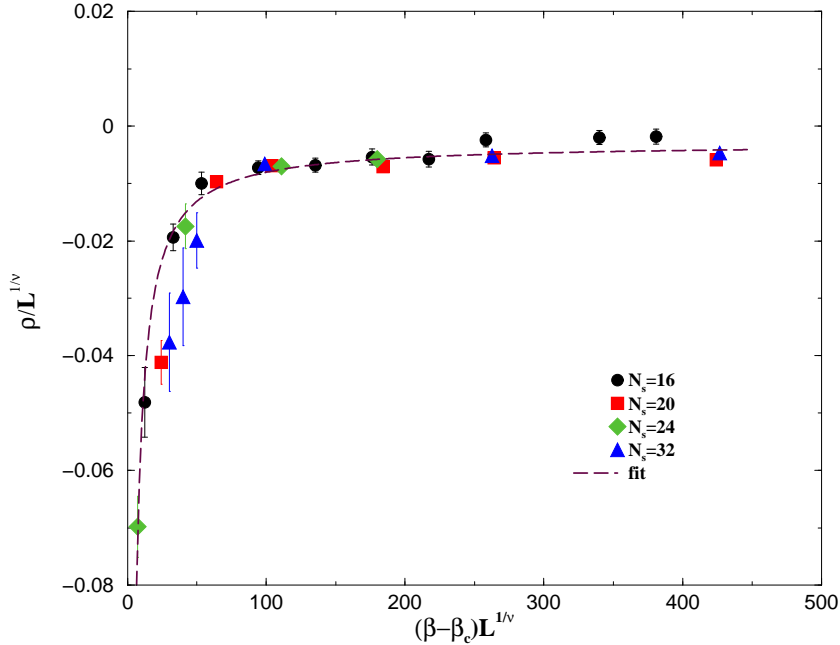, angle=270, width=11cm} 
\end{center}
\caption{Plot of rescaled data.} 
\label{fig:scaling}
\end{figure}

Finite size scaling techniques yield a quantitative description of the
behaviour of $\rho$ in the critical region. The following equation
\begin{eqnarray}
\frac{\rho}{N_s^{1/\nu}} = f\left(N_s^{1/\nu}
\left(\beta_c - \beta\right)\right)
\end{eqnarray}
with $N_s$ being the lattice size, summarises the fact that the rescaled
quantity $\rho/N_s^{1/\nu}$ is a universal function of the scaling
variable
\begin{equation}
x=N_s^{1/\nu}\left(\beta_c - \beta\right)
\end{equation}
For a detailed discussion of the finite size scaling of $\rho$ we
refer to~\cite{ldd95,adg99,ldd00}.  In the $SU(3)$ case, we know that
the transition is first order and there are scaling violations which we
parametrise in the form~\cite{adg99}
\begin{equation}
\label{eq:su3scaling}
\frac{\rho}{N_s^{1/\nu}} = f\left(N_s^{1/\nu}
\left(\beta_c - \beta\right)\right) + \frac{d}{N_s^3}
\end{equation}
Assuming $\nu=1/3$, one can fit the data to Eq.~\ref{eq:su3scaling} to
extract $\beta_c$, $\delta$, $c$ and $d$. We find $\beta_c=5.6924(5)$
in agreement with the result $\beta_c=5.6925$ quoted
in~\cite{boyd95}. A value for $\delta$ can also be extracted. It is
however less stable towards variations of the fitting range and also
depends on the value of $\beta_c$. A conservative estimate, including
a large systematic error, is $\delta=0.51(5)$, which is in agreement
with the value 0.54(4) obtained in~\cite{adg99} from the monopole
condensate. The systematic error is estimated by comparing different
fits to the same data, with a subset of the parameters being kept
fixed, or from the same fit to a subset of the data. The variation in
the fitted parameters is used to estimate the error. The data together
with the fitted curve are displayed in Fig.~\ref{fig:scaling}. 

The data in Fig.~\ref{fig:finitet1} display a difference in the shapes
for the monopole and the vortex peaks on the left of the phase
transition. However, the behaviour inside the peak is described by the
same critical indices.

\section{Discussion}
\label{sect:conc}
Similarly to $SU(2)$, the dual Polyakov line is a good disorder
parameter for deconfinement also in the case of $SU(3)$. It can be
viewed as the dual of the usual Polyakov line, the order
parameter. Once again, its behaviour is similar to that of the
disorder parameter describing condensation of magnetic charges, which
in turn is independent of the choice of the Abelian projection.

We consider these results a step forward in the understanding of the yet
unknown dual fields describing the confined phase. 

\noindent
{\bf Acknowledgements} We thank M. D'Elia, G. Paffuti, M. Teper and A. Kovner
for enlightening discussions. Financial support by the EC TMR Pro\-gram
ERB\-FMRX-CT97-0122 and by MURST is acknowledged. BL is funded by PPARC under
Grant PPA/G/0/1998/00567.

\end{document}